\documentstyle[12pt,epsf]{article}

\begin{document}

\baselineskip 6mm
\renewcommand{\thefootnote}{\fnsymbol{footnote}}

%------------ Hyun Seok's macro's, etc  -----------

\newcommand{\nc}{\newcommand}
\newcommand{\rnc}{\renewcommand}

%\headheight=0truein
%\headsep=0truein
%\topmargin=0truein
%\oddsidemargin=0truein
%\evensidemargin=0truein
%\textheight=9truein
%\textwidth=6.5truein

\rnc{\baselinestretch}{1.24}    % 1.5 spacing btwn text lines
\setlength{\jot}{6pt}       % spacing btwn the rows of an eqnarray
\rnc{\arraystretch}{1.24}       % spacing btwn the rows of a non-eqn array

%%%%%%%%%%%%%%%%%%%%%% Equation Numbering %%%%%%%%%%%%%%%%%%%%%%%
\makeatletter
\rnc{\theequation}{\thesection.\arabic{equation}}
\@addtoreset{equation}{section}
\makeatother

%%%%%%%%%%%%%%%%%%%%%%%%%%%%%%%%%%%%%%%%%%%%%%%%%%%%%%%%%%%%%%%%%
%                               %
%       NEW COMMANDS AND MACROS             %
%                               %
%%%%%%%%%%%%%%%%%%%%%%%%%%%%%%%%%%%%%%%%%%%%%%%%%%%%%%%%%%%%%%%%%

%%%%% Simplify some frequently used LaTeX commands %%%%%

\nc{\be}{\begin{equation}} \nc{\ee}{\end{equation}}
\nc{\bea}{\begin{eqnarray}} \nc{\eea}{\end{eqnarray}}
\nc{\xx}{\nonumber\\} \nc{\ct}{\cite} \nc{\la}{\label}
\nc{\eq}[1]{(\ref{#1})}
\nc{\newcaption}[1]{\centerline{\parbox{6in}{\caption{#1}}}}
\nc{\fig}[3]{

\begin{figure}
\centerline{\epsfxsize=#1\epsfbox{#2.eps}}
\newcaption{#3. \label{#2}}
\end{figure}
}

%%% Caligraphic letters %%%%

\def\CA{{\cal A}}
\def\CC{{\cal C}}
\def\CD{{\cal D}}
\def\CE{{\cal E}}
\def\CF{{\cal F}}
\def\CG{{\cal G}}
\def\CH{{\cal H}}
\def\CK{{\cal K}}
\def\CL{{\cal L}}
\def\CM{{\cal M}}
\def\CN{{\cal N}}
\def\CO{{\cal O}}
\def\CP{{\cal P}}
\def\CS{{\cal S}}
\def\CU{{\cal U}}
\def\CW{{\cal W}}
\def\CY{{\cal Y}}

%%% Double line letters %%%

\def\IR{{\hbox{{\rm I}\kern-.2em\hbox{\rm R}}}}
\def\IB{{\hbox{{\rm I}\kern-.2em\hbox{\rm B}}}}
\def\IN{{\hbox{{\rm I}\kern-.2em\hbox{\rm N}}}}
\def\IC{\,\,{\hbox{{\rm I}\kern-.59em\hbox{\bf C}}}}
\def\IZ{{\hbox{{\rm Z}\kern-.4em\hbox{\rm Z}}}}
\def\IP{{\hbox{{\rm I}\kern-.2em\hbox{\rm P}}}}
\def\IH{{\hbox{{\rm I}\kern-.4em\hbox{\rm H}}}}
\def\ID{{\hbox{{\rm I}\kern-.2em\hbox{\rm D}}}}

%%% Greek letters %%%

\def\a{\alpha}
\def\b{\beta}
\def\ga{\gamma}
\def\d{\delta}
\def\ep{\epsilon}
\def\ph{\phi}
\def\k{\kappa}
\def\l{\lambda}
\def\m{\mu}
\def\n{\nu}
\def\th{\theta}
\def\rh{\rho}
\def\s{\sigma}
\def\t{\tau}
\def\w{\omega}
\def\G{\Gamma}

%%%%% Special Letters

\def\vare{\varepsilon}
\def\bz{\bar{z}}
\def\bw{\bar{w}}

%%%%% Mathematical Symbols

\def\half{\frac{1}{2}}
\def\dint#1#2{\int\limits_{#1}^{#2}}
\def\goto{\rightarrow}
\def\para{\parallel}
\def\brac#1{\langle #1 \rangle}
\def\grad{\nabla}
\def\curl{\nabla\times}
\def\div{\nabla\cdot}
\def\p{\partial}
\def\e{\epsilon_0}

%%%%% Roman aont in math

\def\Tr{{\rm Tr}\,}
\def\det{{\rm dot}}

%%%%% For this paper only

\def\cp{{\bf CP}^n}
\def\s{{\bf S}}
\def\r{{\bf R}}
\def\c{{\bf C}}
\def\z{{\bf Z}}

\begin{titlepage}

%---------------- preprint number ---------------
\hfill\parbox{4cm} {{\tt hep-th/0106269}}

\vspace{25mm}

\begin{center}
%------------------------ title ------------------------
{\Large \bf Topological ${\z}_{N+1}$ Charges on Fuzzy Sphere} \\

\vspace{15mm}
%---------------- authors and addresses ----------------
Chuan-Tsung Chan$^a$\footnote{ctchan@ep.nctu.edu.tw}, Chiang-Mei
Chen$^b$\footnote{cmchen@phys.ntu.edu.tw}, and Hyun Seok
Yang$^b$\footnote{hsyang@phys.ntu.edu.tw}
\\[10mm]
$^a${\sl Department of Electrophysics, National Chiao Tung University \\
Hsinchu 300, Taiwan, R.O.C.}\\
$^b${\sl Department of Physics, National Taiwan University \\
Taipei 106, Taiwan, R.O.C.}

\end{center}

\thispagestyle{empty}

\vskip 20mm

%----------------------- abstract ----------------------
\centerline{\bf ABSTRACT}

\vskip 4mm

\noindent

We study the topological properties of fuzzy sphere. We show that
the topological charge is only defined modulo $N+1$, that is
finite integer quotient $\z_{N+1}$, where $N$ is a cut-off spin
of fuzzy sphere. This periodic structure on topological charges
is shown based on the boson realizations of $SU(2)$ algebra,
Schwinger vs. Holstein-Primakoff. We argue that this result can
have a natural K-theory interpretation and the topological
charges on fuzzy sphere can be classified by the twisted K-theory.
We also outline how solitons on fuzzy sphere can realize D-brane
solitons in the presence of Neveu-Schwarz fivebranes proposed by
Harvey and Moore.

\vspace{2cm}
\end{titlepage}

%-------------------------------------------------------
\baselineskip 7mm
\renewcommand{\thefootnote}{\arabic{footnote}}
\setcounter{footnote}{0}

%%%%%%%%%%%%%%%%%%%%%%%%%%%%%%%%%%%%%%%%%%%%%%%%%%%%%%%%%%%%%%%%%%%%%%
\section{Introduction}
%%%%%%%%%%%%%%%%%%%%%%%%%%%%%%%%%%%%%%%%%%%%%%%%%%%%%%%%%%%%%%%%%%%%%%

The fuzzy sphere is constructed by introducing a cut-off spin $N$
for angular momentum of the spherical harmonics: $\{{\hat Y}_{jm};
j\leq N\}$ \ct{Madore}. Thus the number of independent functions
is $\sum_{j=0}^N (2j+1) = (N+1)^2$. In order for this set of
functions to form a closed algebra, the functions are replaced by
$(N+1) \times (N+1)$ hermitian matrices and then the algebra on
the fuzzy sphere is closed \ct{Hoppe}. Consequently, the algebra
$\CA_N$ on the fuzzy sphere becomes noncommutative matrix algebra
$\mbox{Mat}(N+1)$ which is generated by the fuzzy spherical
harmonics ${\hat Y}_{jm}$, a complete operator basis of
$\CA_N=\mbox{Mat}(N+1)$. The commutative sphere is recovered for
$N \rightarrow \infty$.

The fuzzy sphere we treat here can be obtained from $SU(2)$ group
manifold by the Hopf fibration $\pi:\s^3 \to \s^2$ \ct{CCLY}.
Recently, many insights have been obtained on the geometry of
D-branes on group manifolds $G$ \ct{ars,WZW}. Especially
$G=SU(2)\cong \s^3$ appears as part of Neveu-Schwarz fivebrane
(NS5-brane) geometry and in $AdS_3 \times \s^3 \times M_4$. Group
manifolds, in general, are curved and, in some cases, carry a
non-vanishing NSNS 3-form field $H$. Nevertheless it admits a
full string theory description which rests mainly on conformal
field theories on group manifold, WZW model. The perturbative
analysis based on the $SU(2)$ WZW model shows that the RR charge
of spherical D2-branes is only defined modulo some integer
\ct{WZW}, which is $U(1)$ charge defined on D2-brane world-volume
(fuzzy sphere). This was also confirmed using K-theory
calculation in \ct{K-theory}. Interestingly, the same result was
obtained in the $\cp$ model on fuzzy sphere \ct{CCLY} where the
topological $U(1)$ charge takes values in $\z_{N+1}$.

In spite of these insights from string theory, we think the
topological nature on fuzzy sphere has not been clearly
understood from field theory point of view. Actually we notice the
recent papers \ct{fuzzy} claiming that the $U(1)$ monopole charge,
the first Chern class, is not integer for the fuzzy sphere at
finite cut-off $N$. Thus it should be desirable to study the
topological properties of fuzzy sphere from the field theory
point of view.

This paper is organized as follows. In section 2, in order to set
up our problem, the construction of fuzzy sphere in \ct{CCLY} is
briefly reviewed which is based on Hopf fibration $\pi:\s^3
\rightarrow \s^2$. In section 3, the boson realizations of $SU(2)$
algebra, Schwinger vs. Holstein-Primakoff, and their mapping to
Fock space are studied following the paper \ct{KPTY}. The mapping
method for the $SU(2)$ algebra shows an intriguing periodic
structure between $SU(2)$ representation spaces. This is the
origin of $\z_{N+1}$ topological charges on fuzzy sphere. In
section 4 we argue that the present result nicely fits the recent
twisted K-theory results \ct{K1,KW,K2,K3,BM,K4}. We look into
K-theory interpretation for the results in section 2 and 3. We
also outline how solitons on fuzzy sphere can realize D-brane
solitons in the presence of NS5-branes proposed by Harvey and
Moore \ct{HM}. We address some issues related to our work in
section 5.

%%%%%%%%%%%%%%%%%%%%%%%%%%%%%%%%%%%%%%%%%%%%%%%%%%%%%%%%%%%%%%%%%%%%%%
\section{Fuzzy Sphere from Hopf Fibration}
%%%%%%%%%%%%%%%%%%%%%%%%%%%%%%%%%%%%%%%%%%%%%%%%%%%%%%%%%%%%%%%%%%%%%%

The algebra of fuzzy sphere \ct{Madore,Hoppe} is generated by
$\hat r_a$ satisfying the commutation relations
\begin{equation}\label{NCR}
[ \hat r_a, \, \hat r_b ] = i \alpha\, \epsilon_{abc}\, \hat r_c,
\qquad (a,b,c=1,2,3)
\end{equation}
as well as the following condition for $\hat r_a$:
\begin{equation}\la{R2}
\hat r_a \hat r_a = R^2.
\end{equation}
The noncommutative coordinates of (\ref{NCR}) can be represented
by the generators of the $(N+1)$-dimensional irreducible
representation of $SU(2)$
\begin{equation}\la{rL}
\hat r_a = \alpha \hat L_a,
\end{equation}
where
\begin{equation}
[ \hat L_a, \, \hat L_b ] = i \epsilon_{abc}\, \hat L_c.
\label{su2}
\end{equation}
In the $(N+1)$-dimensional representation space, $\alpha$ and $R$
are related by
\begin{equation}
R^2 = \alpha^2 \frac{N(N+2)}4.
\end{equation}
In the $\alpha \to 0$ limit (or $N \to \infty$ limit) with fixed
$R$, $\hat r_a$ describe commutative sphere:
\begin{equation}
r_1 = R \sin\theta \cos\phi, \quad r_2 = R \sin\theta \sin\phi,
\quad r_3 = R \cos\theta.
\end{equation}

Since $\s^2$ is not parallelizable unlike $\s^3\simeq SU(2)$, the
module of derivations on $\s^2$ is not free \ct{Madore}. If we
enlarge the coordinate space from $\s^2$ to $\s^3$ by the addition
of a $U(1)$ gauge degree of freedom, we can have a free module of
the derivations (acting on $\s^3$). This is a well-known
construction, called the Hopf fibration of $\s^2$. Indeed $\s^3$
can be regarded as a principal fiber bundle with base space $\s^2$
and a $U(1)$ structure group. Equivalently,
\begin{equation}\label{coset}
  \s^2 \simeq SU(2)/U(1),
\end{equation}
where $U(1)$ is the subgroup of $SU(2)$. A complex scalar field on
$\s^2$ can then be identified with a smooth section of this
bundle.

The Hopf fibration $\pi:\s^3 \rightarrow \s^2$ can be generalized
to the noncommutative ${\bf C}^2$ satisfying the relations
\begin{equation} \la{bi-harm}
[a_\a, a_\b] = [a_\a^\dag, a_\b^\dag] = 0, \qquad [a_\a,
a_\b^\dag] = \delta_{\a\b},\qquad (\a,\b,=1,2)
\end{equation}
as follows
\begin{equation} \la{Hopf}
\hat L_a = \frac12 \xi^\dag \sigma_a \xi, \qquad \xi = \left(
\begin{array}{c} a_1 \\ a_2 \end{array} \right),
\la{gens}
\end{equation}
where $\sigma_a$ are the Pauli matrices and $\xi$ is an $SU(2)$
spinor with the normalization $\xi^\dag \xi=N$. Now the $SU(2)$
generators are given by $\c^2$ coordinates
\begin{equation}\la{Ls}
\hat L_1 = \frac12 (a_1 a_2^\dag + a_1^\dag a_2), \quad \hat L_2 =
\frac{i}2 (a_1 a_2^\dag - a_1^\dag a_2), \quad \hat L_3 = \frac12
(a_1^\dag a_1 - a_2^\dag a_2),
\end{equation}
which is, in fact, the Schwinger realization of $SU(2)$ algebra
\ct{Schwinger}. The associated ladder operators are defined as
\begin{equation}
\hat L_+ = \hat L_1 + i \hat L_2 = a_1^\dag a_2, \quad \hat L_- =
\hat L_1 - i \hat L_2 = a_1 a_2^\dag
\end{equation}
and their communication relations are
\begin{equation}\label{L+-3}
[\hat L_+, \hat L_-] = 2 \hat L_3, \qquad [\hat L_3,
\hat L_\pm] = \pm \hat L_\pm.
\end{equation}
Note that the $SU(2)$ generators in \eq{gens} are invariant under
the transformation
\begin{equation}\label{U1}
  \xi \to e^{i \psi} \xi, \quad
  \xi^\dag \to \xi^\dag e^{-i \psi},
\end{equation}
showing that the fiber is $U(1)$.

The $(N+1)$-dimensional irreducible representation of $SU(2)$,
denoted as ${\cal H}_N$, can be given by the following
orthonormal basis \footnote{This representation space ${\cal
H}_N$ satisfies the constraint $(\xi^\dag \xi-N){\cal H}_N=0$,
which is a condition imposed on the space of the irreducible
representation.}
\begin{equation}\label{basis}
|n\rangle =|\frac{N}2, n-\frac{N}2\rangle ={ (a_1^\dag)^{n}
(a_2^\dag)^{N-n}\over \sqrt{n!(N-n)!}}|0\rangle_{12}, \qquad
(n=0,1,\cdots, N),
\end{equation}
where $|j, m\rangle$ is a spherical harmonics and $|0\rangle_{12}$
is the vacuum defined by $a_1|0\rangle_{12}=a_2|0\rangle_{12}=0$.
Let ${\cal A}_N$ be an operator algebra acting on the
$(N+1)$-dimensional Hilbert space ${\cal H}_N$, which can be
identified with the algebra Mat($N+1$) of the complex $(N+1)
\times (N+1)$ matrices. Then the integration over the fuzzy sphere
is given by the trace over ${\cal H}_N$ \footnote{In the
commutative limit, $\Tr$ over matrices is mapped to the
integration over functions as $\Tr \rightarrow \int
\frac{d\Omega}{4\pi}$.}
\begin{equation}
\Tr {\cal O} = \frac1{N+1} \sum_{n=0}^N \langle n | {\cal O} | n
\rangle,
\end{equation}
where ${\cal O} \in {\cal A}_N$.

Let's consider a scalar field $\Phi$, a section of the bundle
\eq{coset}, of the form
\begin{equation}\label{Phi}
\Phi=\sum \Phi_{m_1m_2n_1n_2}{a_1^\dag}{}^{m_1}{a_2^\dag}{}^{m_2}
a_1^{n_1}a_2^{n_2}.
\end{equation}
The above scalar field $\Phi$ can be classified according to the
$U(1)$ gauge transformation \eq{U1} and the set of fields $\Phi$
with definite $U(1)$ charge $k$ will be denoted as $\Phi_k$:
\begin{equation}\label{U1tf}
  \Phi_k \;\;\to \;\;\Phi_k e^{-ik\psi},
\end{equation}
where $k=m_1 +m_2 -n_1 -n_2 \in {\bf Z}$. Indeed the number $k$
labels the equivalence classes (homotopy classes) of $\Phi$ in
\eq{Phi} according to the Hopf fibration \eq{coset}.
\footnote{Note that, for the $\cp$ model in \ct{CCLY}, $\Phi_k :
M \to \cp=SU(n+1)/SU(n) \times U(1)$ and
$\pi_2(\cp)=\pi_1(U(1))=\z$. However, if the manifold $M$ is
noncommutative (in our case $M$ is fuzzy sphere), it will be
shown in section 4 that there is an isomorphism in the homotopy
classes caused by the topological obstruction on the Hopf bundle
\eq{coset}.}

In order to expose topologically nontrivial field configurations,
the Holstein-Primakoff realization of $SU(2)$ algebra \ct{HP} is
more appropriate as pointed out in \ct{CCLY} since they allow us
to directly separate the $U(1)$ symmetry \eq{U1} from $SU(2)$.
Their expressions are given by
\begin{eqnarray}\la{K+-}
  \hat K_+ = a_1^\dag \sqrt{N- a_1^\dag a_1} &+&a_2^\dag \sqrt{N-
a_2^\dag a_2}, \quad \hat K_- = K_+^\dag, \nonumber\\
\hat K_3 = -N &+& (a_1^\dag a_1 + a_2^\dag a_2),
\end{eqnarray}
where
\begin{equation}
\hat K_{\pm} = \hat K_1 \pm i \hat K_2.
\end{equation}
It is straightforward to check the $SU(2)$ algebra
\begin{equation}
[ \hat K_a, \, \hat K_b ] = i \epsilon_{abc}\, \hat K_c,
\label{Ksu2}
\end{equation}
or
\begin{equation}
[\hat K_+, \hat K_-] = 2 \hat K_3, \qquad [\hat K_3, \hat K_\pm] =
\pm \hat K_\pm.
\end{equation}

The derivatives of an operator ${\cal O} \in \CA_N$ are defined by
the adjoint action of $\hat K_a$:
\begin{equation} \label{deriv}
\hat \nabla_a {\cal O} = i[\hat K_a,\, {\cal O}].
\end{equation}
One can see that $\Phi_k$ in \eq{Phi} is an eigenfunction of the
fibration operator $\hat K_3$, namely
\begin{equation}\label{K3=k}
[\hat K_3, \Phi_k]=k \Phi_k.
\end{equation}
So the $SU(2)$ generators \eq{K+-} can be decomposed into the
derivation $\hat K_3$ along the fiber, i.e. the Killing vector
along $U(1) \subset SU(2)$ and the tangent derivations $\hat
K_{\pm}$ to $\s^2 \subset SU(2)$.\footnote{In section 4 we will
more explain the geometrical picture of the Schwinger vs.
Holstein-Primakoff realizations.} We will thus identify the
generators $\hat K_a$ with the derivations acting on the Hopf
bundle \eq{coset}.

%%%%%%%%%%%%%%%%%%%%%%%%%%%%%%%%%%%%%%%%%%%%%%%%%%%%%%%%%%%%%%%%%%%%%%
\section{Schwinger vs. Holstein-Primakoff}
%%%%%%%%%%%%%%%%%%%%%%%%%%%%%%%%%%%%%%%%%%%%%%%%%%%%%%%%%%%%%%%%%%%%%%

In the previous section, we introduced two realizations,
Schwinger and Holstein-Primakoff, of the Lie algebra $SU(2)$ in
terms of boson operators. It was shown by Kuriyama, da
Provid\^{e}ncia, Tsue and Yamamura \ct{KPTY} that the
Holstein-Primakoff representation can be derived from the
Schwinger representation using boson mapping method. (For boson
expansion, boson mapping method, and their applications to nuclear
physics and condensed matter physics, see, for example, two
review articles \ct{KM}.) In this section we will study the boson
realization of $SU(2)$ algebra and their mapping to Fock spaces
to show an intriguing periodic structure of $SU(2)$
representation space, which is the origin of $\z_{N+1}$
topological charges on fuzzy sphere.

To make our discussion as clear as possible, let's start it with
pedagogical way. One can examine the representation matrices of
$SU(2)$ generators ${\hat L}_{\pm},\;{\hat L}_3$ in the space of
states $|j,m\rangle$:
\begin{eqnarray}\label{su2fre}
&&\langle j,m \pm 1|{\hat L}_{\pm}|j,m\rangle=\sqrt{(j\mp m)(j\pm
m+1)},\xx
&&\langle j,m|{\hat L}_3|j,m\rangle=m.
\end{eqnarray}
One can map the sequence of integers or half integers $m$ in
\eq{su2fre} onto a set of non-negative integers $n,\;0\leq n \leq
2j$, by the displacement
\begin{equation}\label{m=n}
  m=-j+n.
\end{equation}
Thus the representations in \eq{su2fre} become, in an obvious
notaion that suppresses the eigenvalue $j$,
\begin{eqnarray}\label{su2bre}
&&\langle n + 1|{\hat L}_+|n\rangle=\sqrt{(n+1)(2j-n)},\xx
&&\langle n - 1|{\hat L}_-|n\rangle=\sqrt{n(2j-n+1)}, \\
&& \langle n|{\hat L}_3|n \rangle=-j+n.\nonumber
\end{eqnarray}
One recognizes immediately that the representations in \eq{su2bre}
can be realized as the matrix elements of the boson operators,
which precisely reduces to the Holstein-Primakoff realization of
$SU(2)$ algebra \ct{HP}
\begin{equation}\label{HP}
{\hat L}^{HP}_+=a^\dag\sqrt{2j-a^\dag a},\quad {\hat
L}^{HP}_-=({\hat L}^{HP}_+)^\dag, \quad {\hat L}^{HP}_3=-j+a^\dag
a,
\end{equation}
where $a,\;a^\dag$ satisfy
\begin{equation}\label{hoa}
  [a,a^\dag]=1
\end{equation}
and the $SU(2)$ operators in \eq{HP} act on a subspace of the
infinite-dimensional boson Fock space with basis
\begin{equation}\label{HPb}
  |n)={1 \over \sqrt{n!}} (a^\dag)^n|0),\qquad n=0,1,2,\cdots.
\end{equation}
In order to preserve the unitary representation of $SU(2)$, the
form of \eq{HP} reminds us that it should be defined in the
subspace $n\leq 2j$, here we call it the physical subspace
according to \ct{KPTY}. However the naive restriction of the
representation space to the physical subspace is not consistent
with the algebra \eq{hoa} since the irreducible representation of
\eq{hoa} requires infinite-dimensional Fock space. Thus one should
pay more careful looking on the mapping between the Schwinger and
the Holstein-Primakoff representations. Here we will follow the
argument in \ct{KPTY}.

For this purpose, let's introduce the following operators
\begin{eqnarray}
\label{B}
  &&\hat B_+ =a^\dag_1 {1\over \sqrt{1+a^\dag_2a_2}} a_2, \quad
  \hat B_-=(\hat B_+)^\dag,\\
\label{C}
  &&\hat C_+ =a^\dag_1 {1\over \sqrt{1+a^\dag_1a_1}} a_2, \quad
  \hat C_-=(\hat C_+)^\dag,
\end{eqnarray}
where $a_\alpha$'s are boson operators in \eq{bi-harm}. Then the
Schwinger generators \eq{Ls} can be expressed independently by
each set of operators in \eq{B} and \eq{C}. The one set is
\begin{equation}\label{HPB}
{\hat L}^B_+=\hat B_+\sqrt{\hat N-\hat B_+ \hat B_-},\quad {\hat
L}^B_-=({\hat L}^B_+)^\dag, \quad {\hat L}^B_3=-{\hat N \over 2}
+\hat B_+ \hat B_-,
\end{equation}
where $\hat N=a^\dag_1a_1+a^\dag_2a_2$ and the other set is
\begin{equation}\label{HPC}
{\hat L}^C_+=\sqrt{\hat N-\hat C_- \hat C_+}\hat C_+,\quad {\hat
L}^C_-=({\hat L}^C_+)^\dag, \quad {\hat L}^C_3={\hat N \over 2} -
\hat C_- \hat C_+.
\end{equation}
Although the above expressions formally resemble the
Holstein-Primakoff realization, they can not be identified with it
since $\hat N$ is an operator instead of a $c$-number and ${\hat
B}_{\pm}$ and ${\hat C}_{\pm}$ are not boson operators such as
\eq{hoa}.

One can easily check that the operators \eq{HPB} and \eq{HPC}
satisfy $SU(2)$ algebra \eq{L+-3}. Also the following relations
can be easily derived
\begin{eqnarray}\label{BC}
   && {\hat N \over 2}({\hat B}_{\pm}|j,m\rangle)=j({\hat
    B}_{\pm}|j,m\rangle), \quad {\hat L}_3({\hat B}_{\pm}|j,m\rangle)=
    (m\pm1)({\hat B}_{\pm}|j,m\rangle),\\
    && {\hat N \over 2}({\hat C}_{\pm}|j,m\rangle)=j({\hat
    C}_{\pm}|j,m\rangle),\quad
    {\hat L}_3({\hat C}_{\pm}|j,m\rangle)=
    (m\pm1)({\hat C}_{\pm}|j,m\rangle), \\
    && [{\hat B}_-, {\hat B}_+]=1-\sum_j (2j+1) |j,j \rangle \langle j,j|, \\
    && [{\hat C}_+, {\hat C}_-]=1-\sum_j (2j+1) |j,-j \rangle \langle
    j,-j|.
\end{eqnarray}
Further, with the help of the above relations, one can deduce the
following properties
\begin{eqnarray}\label{BC+-=0}
 && {\hat B}_+|j,j\rangle =0, \qquad {\hat B}_-|j,-j\rangle =0,\\
 && {\hat C}_+|j,j\rangle =0, \qquad {\hat C}_-|j,-j\rangle =0.
\end{eqnarray}
The above relations imply that there exist lower and upper bounds
with respect to the operation of ${\hat B}_{\pm}$ and ${\hat
C}_{\pm}$.

Now we will consider the boson mapping whose basic idea was
developed by Marumori, Yamamura and Tokunaga \ct{MYT}. Let's
consider the following boson Fock space for \eq{bi-harm}, denoted
as $\CH$:
\begin{equation}\label{Fock}
  \CH=\{|m,n)={1 \over \sqrt{m!n!}}
  (a_1^\dag)^m (a_2^\dag)^n|0\rangle_{12},\; \;m,\,n=0,1,2,\cdots
  \}.
\end{equation}
This Fock space can be decomposed into following way
$$
\CH=\bigoplus_{j \in {{\bf Z} \over 2}}\CH_j,
$$
where $j=(m+n)/2$ and the subspace $\CH_j\;(j=N/2)$ is given by
\eq{basis}. It is well-known that the subspace $\CH_j$ serves the
irreducible representation space of $(2j+1)$ dimensions for the
$SU(2)$ algebra \eq{L+-3}. In other words, the irreducible
representation space $\CH_j$ of spin $j$ is obtained by a
projection from the whole boson Fock space $\CH$, that is
$\CH_j=P_j \CH$, where $P_j$ is the projection operator for the
state with quantum number $j$. This point of view was already
emphasized by Schwinger in his famous paper \ct{Schwinger}.

According to the similar spirit, one can obtain the irreducible
representation of spin $j$ for the Holstein-Primakoff generators
\eq{HP} by a projection from the infinite-dimensional Fock space
\eq{HPb}. Note that we have two kinds of realization for the
Schwinger generators \eq{HPB} and \eq{HPC} in terms of the
operators \eq{B} and \eq{C}, respectively. In general the
representation spaces of \eq{HPB} and \eq{HPC} don't have to be
mapped to the same Fock space. So we will introduce two kinds of
Fock space for the Holstein-Primakoff realization, denoted as
${\cal H}^1$ and ${\cal H}^2$ respectively:
\begin{eqnarray}\label{H1}
{\cal H}^1&=&\{|n)_1
={1\over\sqrt{n!}}(a_1^\dag)^{n}|0\rangle_1,\;
n=0,1,\cdots \},\\
\label{H2} {\cal H}^2&=&\{|m)_2 ={1 \over
\sqrt{m!}}(a_2^\dag)^{m}|0\rangle_2,\; m=0,1,\cdots \},
\end{eqnarray}
where $|0\rangle_{1,2}$ is the vacuum defined by
$a_1|0\rangle_1=a_2|0\rangle_2=0$.

To derive the Holstein-Primakoff representations for \eq{HPB} and
\eq{HPC}, let's introduce the following operators mapping from the
original $SU(2)$ space \eq{basis} to each subspace of \eq{H1} and
\eq{H2}
\begin{eqnarray}
\label{MYT1}
 && U_j^1=\sum_{m=-j}^j |j+m)_1\langle j,m|,\\
\label{MYT2}
 && U_j^2=\sum_{m=-j}^j |j+m)_2\langle j,-m|.
\end{eqnarray}
The properties of $U_j^{1,2}$ are shown in the form
\begin{eqnarray}\label{PU1}
&&  {U^1_j}^\dag U_j^1 =\sum_{m=-j}^j |j,m\rangle\langle j,m|=1,
  \quad  U_j^1 {U^1_j}^\dag =\sum_{m=-j}^j |j+m)_1
  {}_1(j+m|=P_j^1,\\
  \label{PU2}
&&  {U^2_j}^\dag U_j^2 =\sum_{m=-j}^j |j,-m\rangle\langle j,-m|=1,
  \; U_j^2 {U^2_j}^\dag =\sum_{m=-j}^j |j+m)_2
  {}_2(j+m|=P_j^2,
\end{eqnarray}
where $P_j^1$ and $P_j^2$ are the projection operators on each
physical space, that is,
\begin{equation}\label{P}
  {P_j^{1,2}}^\dag=P_j^{1,2},\qquad (P_j^{1,2})^2=P_j^{1,2}.
\end{equation}
From the definitions \eq{MYT1} and \eq{MYT2}, the following
relations are derived
\begin{eqnarray}
\label{sph-bf}
  && U_j^1|j,m \rangle = |j+m)_1, \quad U_j^2|j,-m \rangle = |j+m)_2,\\
  && {U_j^1}^\dag |j+m)_1=|j,m\rangle,\quad {U_j^1}^\dag |2j+k)_1=0,\xx
\label{bf-sph}
  && {U_j^2}^\dag |j+m)_2=|j,-m\rangle,\quad {U_j^2}^\dag |2j+k)_2=0,\quad
  (k=1,2,\cdots).
\end{eqnarray}
This shows that the original $SU(2)$ space \eq{basis} is in
one-to-one correspondence with the subspace
$\CH_j^{1,2}=\{|n)_{1,2},\; n=0,1,\cdots,2j\}$.

With the use of $U_j^{1,2}$, the operators \eq{B} and \eq{C} are
transformed to the following forms in the corresponding Fock space
\eq{H1} and \eq{H2}, respectively,
\begin{eqnarray}\label{tB}
&& B^b_+ =U^1_j \hat B_+ {U^1_j}^\dag=a_1^\dag
\sum_{n=0}^{2j-1}|n)_1{}_1(n|=a_1^\dag(1-
\sum_{k=0}^{\infty}|2j+k)_1{}_1(2j+k|), \\
&& B^b_-=(B^b_+)^\dag,\xx
\label{tC}
&& C^b_- =U^2_j \hat C_+
{U^2_j}^\dag=a_2 \sum_{n=0}^{2j}|n)_2{}_2(n|=a_2(1-
\sum_{k=1}^{\infty}|2j+k)_2{}_2(2j+k|),\\
&& C^b_+=(C^b_-)^\dag.\nonumber
\end{eqnarray}
Note that the operator $\hat C_+\,(\hat C_-)$ is realized as an
annihilation (creation) operator $a_2\,(a_2^\dag)$ in the space
\eq{H2}, so we have flipped the notation in \eq{tC}, $+
\leftrightarrow -$. The above operators obey the commutation
relation
\begin{eqnarray}\label{BBCC}
&& [B^b_-, B^b_+ ]=P_j^1-(2j+1)|2j)_1{}_1(2j|,\\
&& [C^b_-, C^b_+ ]=P_j^2-(2j+1)|2j)_2{}_2(2j|.
\end{eqnarray}
Also, using the results in \eq{tB} and \eq{tC}, one can easily
check that the Schwinger operators in \eq{HPB} and \eq{HPC} are
transformed to the following forms
\begin{eqnarray}\label{S1}
&&{\hat S}^1_+=P_j^1\Bigl(a_1^\dag\sqrt{2j-a_1^\dag
a_1}\Bigr)P^1_j,\quad {\hat S}^1_-=({\hat S}^1_+)^\dag, \quad
{\hat
S}^1_3=P_j^1(-j+a_1^\dag a_1)P_j^1,\\
\label{S2} &&{\hat S}^2_+=P_j^2\Bigl(a_2^\dag\sqrt{2j-a_2^\dag
a_2}\Bigr)P^2_j,\quad {\hat S}^2_-=({\hat S}^2_+)^\dag, \quad
{\hat S}^2_3=P_j^2(-j+a_2^\dag a_2)P_j^2.
\end{eqnarray}
In \eq{S2} we have also flipped the notation, $+ \leftrightarrow
-$, and thus we have defined ${\hat S}^2_3=-U_j^2 {\hat
L}^C_3{U_j^2}^\dag$.

Using the transformations \eq{MYT1} and \eq{MYT2}, we now obtained
two sets of Holstein-Primakoff generators which are exactly the
same that \eq{HP} in the physical subspaces $\CH_j^1=P_j^1\CH^1$
and $\CH_j^2=P_j^2\CH^2$. However, as pointed out in \ct{KPTY},
the mapping from the original $SU(2)$ space \eq{basis} to the
physical space $\CH_j^1$ or $\CH_j^2$ is not unique. There exist
infinitely many physical spaces for the $SU(2)$ space with spin
$j$ in the whole Fock space $\CH^1$ or $\CH^2$. The physical
subspace $\CH_j^1$ or $\CH_j^2$ is classified by the equivalence
relation given by
\begin{equation}\label{eqclass}
  |\alpha; j+m)_{1,2} \, \sim \,|j+m)_{1,2},
\end{equation}
where $|\alpha; j+m)_{1,2}=|(2j+1)\alpha+(j+m))_{1,2}$ with a
positive integer $\alpha=0,1,2,\cdots.$ The equivalence classes
are characterized by the rank of the projection operators
$P_j^{1,2}$, i.e. $(2j+1)$. Actually, these equivalence classes
define K-group of a (compact) operator algebra in the Hilbert
space $\CH=\CH^1 \bigotimes \CH^2$.

Here we will not repeat the analysis in order to show the
equivalence relation \eq{eqclass} since it was already done in
\ct{KPTY} and it is a straightforward (and simple) algebra.
Instead, let's briefly summarize the methodology. It starts with
the following correspondence,
\begin{equation}\label{correspondence}
  |j,m\rangle \,\leftrightarrow \, |\alpha; j+m)_{1,2},\qquad
  \alpha=0,1,2,\cdots
\end{equation}
according to \eq{sph-bf} and \eq{bf-sph}. (The transformations
\eq{MYT1} and \eq{MYT2} correspond to selecting one
representative class with $\alpha=0$ as a physical subspace.)
Using this correspondence, one can find the operators
${\widetilde{B}}_{\pm},\; {\widetilde{C}}_{\pm}$ acting on the
whole Fock space $|\alpha; j+m)_{1,2} \; (\alpha=0,1,\cdots)$,
which correspond to $B^b_{\pm},\; C^b_{\pm}$, and the $SU(2)$
operators ${\widetilde{S}}^{1,2}_{\pm,3}$ in terms of
${\widetilde{B}}_{\pm},\; {\widetilde{C}}_{\pm}$, respectively.
Then one can check that the tilde operators are well-defined in
the {\it whole} Fock space and the basis $|\alpha; j+m)_{1,2}$
for each $\alpha$ provides the {\it same} $SU(2)$ representation
for ${\widetilde{S}}^{1,2}_{\pm,3}$, where the states with
different $\alpha$ can not be connected by the operations of
${\widetilde{S}}^{1,2}_{\pm}$, i.e. completely disconnected. Thus
one can choose a physical subspace, e.g. $\alpha=0$, as the
representation space for the Holstein-Primakoff realization.

Note that the $SU(2)$ operators ${\hat K}$ in \eq{K+-} are the sum
of two spin operators ${\hat S}_{1},\;{\hat S}_{2}$,
Eqs.\eq{S1}-\eq{S2}, in the physical subspaces $\CH^1_N$ and
$\CH_N^2 \;(j=N/2)$:
\begin{equation}\label{K}
  {\hat K}_a={\hat S}_{1a}+{\hat S}_{2a},
\end{equation}
where
\begin{equation}\label{SS}
  [{\hat S}_{1a},{\hat S}_{1b}]=i\epsilon_{abc}{\hat S}_{1c},
  \quad [{\hat S}_{2a},{\hat S}_{2b}]=i\epsilon_{abc}{\hat
  S}_{2c}, \quad [{\hat S}_{1a},{\hat S}_{2b}]=0.
\end{equation}
Then the basis \eq{basis} is a tensor product of ${\cal H}_N^1$
and ${\cal H}_N^2$ and can be expanded in the basis of total spin
operator, that is,
\begin{equation}\label{decH}
{\cal H}_N^1\bigotimes{\cal H}_N^2=\bigoplus_{J=0}^{N}{\cal
H}^{(J)}
\end{equation}
and
\begin{equation}\label{H}
\bigoplus_J{\cal H}^{(J)}=\{|J, 0\rangle, \; J=0,1,\cdots, N\},
\end{equation}
where the spherical harmonics $|J, M\rangle$ is a spin-$J$
representation of the operator ${\hat K}_a$:
\begin{equation}\label{totalbasis}
{\hat K}^2_a|J, M\rangle=J(J+1)|J, M\rangle, \;\;\; {\hat K}_3|J,
M\rangle=M |J, M\rangle.
\end{equation}
Thus the states in \eq{H} can serve as the $(N+1)$-dimensional
bases of ${\hat K}_a$ \ct{CCLY}.

In our previous paper \ct{CCLY}, we showed that the topological
charge on fuzzy sphere is defined by the number of flux units
passing through the fuzzy sphere and it is given by an eigenvalue
of $\hat{K}_3$. It should be emphasized, however, that this is a
general property of the Hopf bundle \eq{coset}. As seen above, the
representation space or the physical space of Holstein-Primakoff
algebra has the periodic structure \eq{eqclass} in the whole Fock
space whose periodicity is $2j+1$ or $N+1$. Thus it means that
the eigenvalue of $\hat{K}_3$, so the topological charge, is also
only defined modulo $N+1$, that is finite integer quotient
$\z_{N+1}$. This is the origin of the $\z_{N+1}$ topological
charge on fuzzy sphere.

In the next section, we will discuss why the present result nicely
fits the twisted K-theory. We will also discuss the geometrical
aspects of Schwinger vs. Holstein-Primakoff realization of
$SU(2)$ algebra. This will provide a picture on how solitons on
fuzzy sphere can realize D-brane solitons in the presence of
NS5-branes proposed by Harvey and Moore \ct{HM}. However we will
not try any rigorous proofs on the K-theory approach, but just
outline the K-theory interpretation about the results in section
2 and 3. Nevertheless we hope some elaborate aspects of twisted
K-theory would be obvious from our ``physical" setup.

%%%%%%%%%%%%%%%%%%%%%%%%%%%%%%%%%%%%%%%%%%%%%%%%%%%%%%%%%%%%%%%%%%%%%%
\section{$K_H^*(SU(2))$ and D-brane Solitons}
%%%%%%%%%%%%%%%%%%%%%%%%%%%%%%%%%%%%%%%%%%%%%%%%%%%%%%%%%%%%%%%%%%%%%%

K-theory provides a mathematical framework classifying
Ramond-Ramond charges and fields \ct{K1,KW,K2,K3,BM,K4}. The
K-theory of commutative space $X$, $K(X)$, can be generalized to
that of noncommutative space if $C(X)$, a commutative ring of
continuous functions on $X$, is replaced by a noncommutative
$C^*$-algebra \ct{K-books}. In this case, the topological K-theory
turns out to have a natural link with the theory of operators in
Hilbert space $\CH$ which is a representation space (of infinite
dimension) of $C^*$-algebra. So we need to consider bundles of
Hilbert spaces and the `large' group $U(\CH)$ of all unitary
operators in Hilbert space. (Recently it was proposed \ct{Harvey}
that the gauge group of noncommutative field theory is
$U_{\mbox{cpt}}(\CH)$ defined by taking the maximal completion of
finite rank operators with respects to $L^p$-norm.) We have an
exact sequence of groups $(n\to \infty)$:
\begin{equation}\label{exactseq}
  0\;\to\; U(1)\;\to\;U(n)\;\to\;PU(n)\;\to\;0.
\end{equation}
This sequence implies that there could be an obstruction $H \in
H^3(X, {\bf Z})$ to lifting a projective bundle to a vector
bundle.

If $H \in H^3(X, {\bf Z})$ is a torsion class, i.e. $n \cdot
H=0$, the construction of K-theory is following \ct{KW,K3,BM,K4}.
Given a vector bundle $V$ on $X$ we can form the projective
principal bundle $P(V)$ with structure group $PU(n)$. For any
vector space $V$, $\mbox{End}\,V=V \otimes V^*$, endomorphism of
$V$, depends only on $P(V)$. Hence, given a projective bundle $P$
over $X$ we can define the associated bundle $\CE(P)$ of
endomorphism (matrix) algebras. The sections of $\CE(P)$ form a
noncommutative $C^*$-algebra and one can define its K-group by
using finitely generated projective modules. This K-group defines
the twisted K-group $K_H^*(X)$ and it depends only on its
obstruction class $H \in H^3(X, {\bf Z})$.

If $H \in H^3(X, {\bf Z})$ is not a torsion element, then we
somehow need to take the limit $n \to \infty$, that is, we need
to consider bundles of Hilbert spaces over $X$ \ct{BM,K4}. This
leads us to the study of principal bundle $P_H$ over $X$ with
fiber $PU(\CH)=U(\CH)/U(1)$ where $\CH$ is an infinite
dimensional, separable, Hilbert space. If the $C^*$-algebra of
compact operators on $\CH$ is denoted by $\CK$, one can identify
$PU(\CH)=Aut(\CK)$, automorphism of $\CK$. Let $\CE_H$ be the
associated principal bundle $P_H \times_G \CK$ where $G=Aut(\CK)$
also acts on $\CK$ by conjugation. This bundle is also completely
classified by $H^3(X,{\bf Z})$. Then the twisted K-theory
$K_H^*(X)$ for non-torsion $H \in H^3(X, {\bf Z})$ is defined as
the homotopy classes of the $C^*$-algebra of continuous sections
of $\CE_H$.

As mentioned in the Introduction, noncommutative $C^*$-algebra on
fuzzy sphere is given by the universal enveloping algebra $\CU$ of
$SU(2)$, which is generated by all completely symmetrized
polynomials in $SU(2)$ generators and whose irreducible bases are
given by fuzzy spherical harmonics $\hat Y_{jm}$
\ct{Madore,Hoppe}. As a representation of the noncommutative
$C^*$-algebra, we can take the endomorphism algebra for a vector
space $\CH_j$, $\mbox{End}\,\CH_j =\CH_j \otimes \CH^*_j$, which
is a finite dimensional matrix algebra. However, in order to
construct the $C^*$-algebra for fuzzy sphere, it is more natural
to introduce an infinite-dimensional algebra representing, e.g.
noncommutative $\c^2$ such as \eq{bi-harm}, although its twisted
K-theory has a finite dimensional description (as illustrated in
section 3). The necessity of going to infinite dimensional algebra
bundles in order to incorporate nontorsion classes was already
pointed out in \ct{BM} and was interpreted as going off-shell in
\ct{KW}.

We have a fixed infinite dimensional, separable, Hilbert space
$\CH={\cal H}^1 \bigotimes {\cal H}^2$ given by \eq{Fock} that is
an irreducible representation space of noncommutative $\c^2$.
Using the algebra $B(\CH)$ of all bounded operators on $\CH$ and
Fredholm operators, i.e. operators in $B(\CH)$ with finite
dimensional kernel and cokernel, we can construct the Hilbert
space bundle $\CE_H$ with fiber $\CK$ and transition function
$G=Aut(\CK)$. Then the scalar field $\Phi$ defined in \eq{Phi}
corresponds to a continuous section of the algebra bundle
$\CE_H$. According to Atiyah and Segal, in order to define the
twisted K-theory $K_H^*(SU(2))$, we have to study the homotopy
classes of the section $\Phi$ \ct{K4}.

The exact sequence \eq{exactseq} implies that there can an
obstruction to go from a projective bundle $P_H$ to a unitary
bundle with fiber $U(\CH)$ if the third cohomology group of base
manifold $X,\;H^3(X, \z)$, does not vanish. It was known
\ct{BM,K4} that this obstruction class $\delta(P_H)=[H] \in H^3(X,
\z)$, so called Dixmier-Douady class, completely classifies the
isomorphism classes of the bundles $P_H$ and $\CE_H$. For example,
$\omega^3 ={1\over 3!} \theta \wedge \theta
 \wedge \theta \in H^3(SU(2), \z)$ in Lie algebra cohomology
where $\theta$ is the left-invariant Cartan-Maurer 1-form on
$SU(2)$, and NS 3-form field strength $H \in H^3(\s^3, \z)$ in
NS5-brane geometry. Actually, these classes are intimately
related in the description of D-branes on $SU(2)$ manifold in
terms of WZW model. In this context, these classes appear as the
obstruction to defining a line bundle $\CL$ (locally defined) on a
submanifold $D \subset SU(2)$ as follows \ct{WZW}
\begin{equation}\label{lineb}
  \int_D \omega - \int_Z H = \int_D F,
\end{equation}
where $\omega$ is a 2-form on $D$ such that $H=d\omega$ and $Z$
is a 3-cycle in $SU(2)$ such that $\partial Z =D$. The right-hand
side of \eq{lineb} corresponds to D0-brane charges given by the
flux of $U(1)$ gauge field. The left-hand side of \eq{lineb}
depends not only on $D$ but also on $Z$, whereas the right-hand
side depends only on $D$. Thus if we define the D0-brane charge
in another 3-cycle $Z'$ but $\partial Z'=D$, the difference in
the charges computed by $Z$ and $Z'$ reduces to the integral of
$H/2\pi$ on the 3-cycle $Z-Z'$ of $SU(2)$
\begin{equation}\label{H3}
{1\over 2\pi} \int_{Z-Z'} H \in {\bf Z}.
\end{equation}
Since physically the D0-brane charge should not depend on $Z$, we
must define it modulo some integer determined by the obstruction
class $H \in H^3(SU(2), \z)$ \ct{WZW}.

Similar thing happens for our case. In section 2 we constructed
the fuzzy sphere from the Hopf fibration $\pi:\s^3 \to \s^2$,
Eq.\eq{Hopf}. For commutative space, the topological class of the
Hopf map is given by the Hopf invariant
\begin{equation}\label{Hopfmap1}
H(\pi)={1\over 4 \pi^2} \int_{\s^3} AdA
\end{equation}
where $A$ is the $U(1)$ gauge potential defined by the spinor
$\xi$ like \eq{Hopf}
\begin{equation}\label{HA}
  A=-i \xi^\dag d\xi.
\end{equation}
It is known \ct{Bott} that the Hopf invariant depends only on the
homotopy class of the map $\pi$, i.e. $\pi_3(\s^2) \in \z$. Also
it has a neat geometric interpretation: The Hopf invariant of the
map is equal to the linking number of the inverse images of any
two distinct points of $\s^2$.

Now a natural question is what happens for the Hopf invariant in
noncommutative space, e.g., defined by the operator spinor $\xi$
in \eq{Hopf}. The Hopf ``invariant" can be naturally generalized
to a noncommutative analogue of the Chern-Simons form
\begin{equation}\label{Hopfmap2}
H(\pi)={1\over 4 \pi^2} \int_{\s^3} \Bigl(AF-{i\over 3}A^3\Bigr),
\end{equation}
where $F=dA+iA^2+*A$ is gauge-covariant field strength \ct{CCLY}
and $*$ means Hodge dual. However, unlike the commutative case
\eq{Hopfmap1}, the above Hopf ``invariant" is gauge variant under
the large gauge transformation $g \in U(1)$, namely
\begin{equation}\label{U(1)}
A \to g^\dag A g - i g^\dag d g.
\end{equation}
Here the local gauge group $g$ can be understood as unitary
operators on $\CH$. According to \ct{Harvey} and \ct{Flag}, we
will identify the local gauge transformation $g$ with the unitary
operator $U_{\mbox{cpt}}(\CH)$ of the form $g(\CH)={\bf 1} + K$
with $K$ a compact operator, that is $g: \s^3 \rightarrow
U_{\mbox{cpt}}(\CH)$. The change under the local transformation
\eq{U(1)} is then
\begin{equation}\label{dH}
  \Delta H(\pi)={1\over 24 \pi^2} \int_{\s^3}\eta \wedge \eta \wedge
 \eta,
\end{equation}
where $\eta=g^\dag d g$ is the Cartan-Maurer 1-form on
$U_{\mbox{cpt}}(\CH)$. Note that $g^*\eta=\theta$. The quantity
\eq{dH} is the ``winding number" for the mapping $\s^3 \to
U_{\mbox{cpt}}(\CH)$ whose homotopy class is given by
$\pi_3(U_{\mbox{cpt}}(\CH))=\z$. (This is essentially the same as
the quantization condition on the level of Chern-Simons theory in
\ct{NP}. See, also, \ct{Kra} for a related argument on the
quantization condition.)

As discussed before, the $C^*$-algebras on the fuzzy sphere are
defined through the Hopf map \eq{Hopf}. But the homotopy class of
the Hopf map should be defined only modulo some integer class
determined by \eq{dH} because it is generated by local gauge
transformation as just shown above. This equivalence class should
be an isomorphism class of the $C^*$-algebra including the
algebra bundle $\CE_H$ and its sections on the fuzzy sphere.
Consequently the isomorphism class of the algebra bundle $\CE_H$
and its sections are determined by the obstruction class which is
given by $\omega^3 ={1\over 3!} \eta \wedge \eta \wedge \eta$ and
the isomorphism class is uniquely defined (up to Murray-von
Neumann equivalence) for a given obstruction class. We see that
in our case the above isomorphism has been realized as the
periodic structure \eq{eqclass} acting on $\CH$.

This is very similar to the situation of $SU(2)$ WZW model since
here D-brane world-volumes are defined by conjugacy classes of
$SU(2)$, characterized by elements of maximal torus \ct{ars,WZW}.
This defines $SU(2)/U(1)$, so Hopf fibration, which is isomorphic
to the (fuzzy) sphere except as degenerate points at $\{e,-e\}$.
But the existence of nontrivial NS3-form field $H \in H^3(\s^3,
\z)$ gives the obstruction \eq{H3} upon the construction of line
bundle $\CL$ over $SU(2)/U(1)$. Therefore for nontrivial $H$ the
D0-brane charge, which is the first Chern class of (locally
defined) line bundle $\CL$, takes values in an integer quotient,
for instance, $\z_{N+1}$ \ct{K-theory,CCLY}.

In section 3, using the relation between Schwinger and
Holstein-Primakoff realizations of $SU(2)$ algebra, it has been
shown that the representation, so their $C^*$-algebra $\CA_N$, of
fuzzy sphere is defined by the equivalence class \eq{eqclass} in
the whole Fock space $\CH$. Thus the algebra $\CA_N$ itself is
realized in the finite dimensional space $\CH_N$ rather than
$\CH$ determined by the rank of the projection operators such as
\eq{PU1} and \eq{PU2}. This is just the property of $K^*_H(X,\z)$
\ct{BM}. Although the underlying $C^*$-algebra is
infinite-dimensional, its twisted K-theory has finite-dimensional
description since the range of the projection operators in $\CK$
is a finite dimensional subspace $\CH_N \subset \CH$. So we see
the topological properties of fuzzy sphere nicely fit the twisted
K-theory results \ct{BM,K4}.

The $SU(2)\cong \s^3$ group manifold appears as part of NS5-brane
geometry and in $AdS_3 \times \s^3 \times M_4$. In order to
speculate how solitons on fuzzy sphere can realize D-brane
solitons in the presence of $H \in H^3(X, {\bf Z})$ proposed by
Harvey and Moore \ct{HM}, let's consider $N$ coincident type IIA
NS5-branes. (For the dynamics of D-branes in the geometry of
NS5-branes, see also \ct{NS5}). We will take $X=\r^{5,1} \times
\r^4$ with $\r^{5,1}$ to be the world-volume of NS5-branes. Let's
decompose the transverse space $\r^4$ into a radial coordinate
$r\in \r^+$ and $u=(\theta, \phi,\psi) \in \s^3 $:
\begin{equation}\label{r4}
  \r^4 \cong \s^3 \times \r^+.
\end{equation}
The quantized 5-brane charge $N$ is then given by
\begin{equation}\label{k}
  N=\int_{\s^3}H,
\end{equation}
where $H \in H^3(X, {\bf Z})$ is the NS 3-form field strength.
Noting that $\s^2 \simeq \s^3/U(1)$ and $\s^1 \simeq \r/\z$, the
space $\s^3 \times \r$ in \eq{r4} can be regarded as a principal
fiber bundle over $\s^2 \times \s^1$ with fiber $U(1) \times \z$.
The $U(1) \times \z$ transition functions act on $\s^3 \times \r$
by
\begin{eqnarray}\label{tranfn}
  (u, r) &\sim & (e^{i\chi}u, r), \xx
(u, r) &\sim & (u, r+1),
\end{eqnarray}
where the first line is the left $U(1)$ action on $u \in \s^3$ and
the second line is the $\z$ action on $r \in \r^+$. Then $X=\s^2
\times \s^1$ allows an explicit realization of the $PU(\CH)$
bundles over $X$ for $[H] \in H^3(\s^2 \times \s^1, \z)=\z$ by
embedding the $U(1) \times \z$ transition functions into
$PU(\CH)$ \ct{Bryl}. \footnote{As pointed out in \ct{Harvey,Flag},
$U_{\mbox{cpt}}(\CH)$ rather than $PU(\CH)$ could be more
appropriate structure group for nontrivial Hilbert bundles as
illustrated in \eq{dH}. Nevertheless we will not distinguish them
since it is not essential for the following argument.} (A similar
but more elaborate construction for $X=\s^3$ may be extracted
from \ct{RR}.)

According to the construction in section 2, we will take $\r^4$ to
be the noncommutative space represented by \eq{bi-harm}. In terms
of the Hopf map \eq{Hopf}, the noncommutative $\c^2\cong \r^4$
algebra in \eq{bi-harm} is transformed to $U(2)$ algebra whose
generators are given by $\{\hat L_a, \hat N=\xi^\dag\xi\}$. Note
that $[\hat L_a, \hat N]=0$. From \eq{K3=k}, one can see that the
position operator $\hat \psi$ of $U(1)$, the fiber in the Hopf
fibration $\s^3 \to \s^2$, and $\hat N$ the integrally quantized
radius are conjugate variables, that is, $[\hat \psi, \hat N]=i$.
Using these operators we can form a representation of $U(1)
\times \z$ in $PU(\CH)$ via
\begin{equation}\label{puh}
  (e^{i\chi}, n) \to e^{in\hat \psi}e^{i\chi\hat N}.
\end{equation}
Since $e^{in\hat \psi}$ and $e^{i\chi\hat N}$ commute up to the
phase $e^{in\chi}$, \eq{puh} is indeed a representation of $U(1)
\times \z$ transition functions in $PU(\CH)$.

In order to define $PU(\CH)$ connection $\omega$, we will
pull-back the volume-form $\Omega$ on the sphere into $\s^3$, i.e.
\begin{equation}\label{pull-back}
  \pi^* \Omega = d\omega
\end{equation}
where 1-form $\omega$ can be locally defined by separating $\s^3$
into two hemispheres $U_{\pm}$ with common boundary $\s^2$. Then
we can define Lie algebra ($\hat K_3$) valued $PU(\CH)$ connection
as \ct{HM}
\begin{equation}\label{connection}
\omega(U_{\pm}, r)=i\Bigl(d\psi \pm (1\mp \cos \theta) d\phi
\Bigr) (\hat N -r {\bf 1}).
\end{equation}
The above connection has the property
\begin{equation}\label{omegapd}
\omega(U_{\pm}, r+k)=e^{ik\hat \psi} \omega(U_{\pm}, r)
e^{-ik\hat \psi}.
\end{equation}
From this connection over $\s^2 \times \s^1$, the globally defined
curvature 3-form $K=-{i\over 2} \sin\theta d\theta d\phi dr$ with
$\int_{\s^2 \times \s^1} K/2\pi i =1$, can be identified with the
Dixmier-Douady class for twisted $PU(\CH)$ bundle \ct{Bryl,HM}.

Note that in our case the eigenvalue $N$, the quantized radius, of
$U(1)$ operator $\hat N$ specifies the cut-off spin of fuzzy
sphere, so it determines the rank of the projection operators in
$\CH$. And the projection operators selecting the physical
subspace have the isometry \eq{eqclass}, the periodic structure
for $SU(2)$ representation space. Since the connection
\eq{connection} can be regarded as that of a line bundle (locally
defined) over $SU(2)/U(1)$, the periodic structure \eq{omegapd} in
the connection $\omega$ can be related to that of \eq{eqclass}
for the following reason. By transgression \ct{Loop}, the classes
in $H^3(G, \z)$ are in one-to-one correspondence with classes in
$H^2(LG,\z)$, where $LG$ is the loop group of $G$. According to
Borel-Weil theory, the irreducible representation of simple Lie
group $G$ is given by holomorphic sections of a line bundle
$\CL_\lambda$ over $G/T$ for a representation $\lambda: T \to
\s^1$. And the first Chern class in $H^2(LG,\z)$ competely
describes the topological type of the line bundle. So it seems
natural that the periodic structure of the connection of
$\CL_\lambda$ is intimately related to that of representation
space, although its rigorous formulation is certainly required
(see the last paper in \ct{WZW}).

In view of the above, we expect that solitons on fuzzy sphere, as
an example, the BPS solitons constructed in \ct{CCLY}, can
naturally realize D-brane solitons in the presence of NS5-branes
proposed by Harvey and Moore. Of course, it is not obvious how to
construct the low-energy effective action describing this system
under the NS5-brane background as well as tachyon background.
However we believe that the essential feature outlined in this
paper can be applied even in this case because both the solitons
on fuzzy sphere and the D-brane solitons in the NS5-brane
background have the same kind of topological charge defined by
the twisted K-theory $K_H^*(X,\z)$.

%%%%%%%%%%%%%%%%%%%%%%%%%%%%%%%%%%%%%%%%%%%%%%%%%%%%%%%%%%%%%%%%%%%%%%
\section{Discussion}
%%%%%%%%%%%%%%%%%%%%%%%%%%%%%%%%%%%%%%%%%%%%%%%%%%%%%%%%%%%%%%%%%%%%%%

We studied the topological properties of fuzzy sphere based on the
boson realizations of $SU(2)$ algebra such as Schwinger and
Holstein-Primakoff. We argued that these results can have a
natural K-theory interpretation and the topological charges on
fuzzy sphere can be classified by the twisted K-theory. Although
the present arguments are consistent with string theory results,
more rigorous formulation is certainly required and it may expose
a deep connection between the representation theory of loop group
$LG$ and the twisted K-theory, for instance, announced in
\ct{Freed}.

In order to study the dynamics of D2-branes wrapping on a 2-cycle
(fuzzy sphere) in $SU(2)$, we have to study
Maxwell-Chern-Simons-Higgs theory or matrix theory on fuzzy sphere
since it is related to the world-volume theory of spherical
D2-branes formed by the bound state of $k$ D0-branes \ct{ars}. As
argued in section 4, since the twisted K-theory topologically
classifies the algebra bundles on fuzzy sphere, we expect that
this theory also enjoys the same properties as the $\cp$ model in
\ct{CCLY} even though the dynamics of gauge fields is considered.
This will help to understand the Harvey and Moore's proposal more
closely. We hope to address this problem soon.

Anomalies in gauge theories can be given a topological
interpretation associated with gauge bundles. In our previous
paper \ct{CCLY}, we showed that the zero modes of Dirac operator
on fuzzy sphere are also given by the $\z_{N+1}$ topological
charge of background solitons. So it is expected, in this case,
the anomalies on fuzzy sphere are deformed by the noncommutativity
and can be related to the twisted K-theory as well \ct{LY}.

There is an another Hopf fibration $\pi: \s^7 \to \s^4$ which is
related to instanton bundles on (fuzzy) $\s^4$. What happens when
we willing to extend the present analyses to this case ? $\s^7$
is also parallelizable manifold which can be defined by
octonions, so we can have a free module of derivations acting on
$\s^7$. In this case, however, the fiber is $SU(2)$, non-Abelian,
and, unfortunately, the octonion algebra is non-associative
\ct{octo}. Thus the problem becomes much more complicated.
Nevertheless this problem should be important since it is related
to the instanton solutions, the summit of gauge theory. While the
(noncommutative) Hopf fibration $\pi: \s^3 \to \s^2$ has been
appeared in the geometry of string theory, the Hopf fibration
$\pi: \s^7 \to \s^4$ may be related to M-theory backgrounds since
$\s^7 /\s^4$ appears in the geometry of M2/M5-branes in eleven
dimensions, so may be the level three theory, e.g. elliptic
cohomology, in the hierachy by Witten \ct{KW}.

\section*{Acknowledgments}
We are grateful to Pei-Ming Ho and Feng-Li Lin for helpful
discussions. Two of us are supported by NSC (CTC:
NCS89-2112-M-009-006 and HSY: NCS89-2811-M-002-0095). CMC is
supported by the CosPA project of the Ministry of Education,
Taiwan. We also acknowledge NCTS as well as CTP at Taida for
partial support.

%\newpage

%%%%%%%%%%%%%%%%% Journal Macros %%%%%%%%%%%%%%%%%%%%%%%%%%%

\nc{\np}[3]{Nucl. Phys. {\bf B#1} (#2) #3}

\nc{\pl}[3]{Phys. Lett. {\bf B#1} (#2) #3}

\nc{\prl}[3]{Phys. Rev. Lett. {\bf #1} (#2) #3}

\nc{\prd}[3]{Phys. Rev. {\bf D#1} (#2) #3}

\nc{\ap}[3]{Ann. Phys. {\bf #1} (#2) #3}

\nc{\prep}[3]{Phys. Rep. {\bf #1} (#2) #3}

\nc{\ptp}[3]{Prog. Theor. Phys. {\bf #1} (#2) #3}

\nc{\rmp}[3]{Rev. Mod. Phys. {\bf #1} (#2) #3}

\nc{\cmp}[3]{Comm. Math. Phys. {\bf #1} (#2) #3}

\nc{\mpl}[3]{Mod. Phys. Lett. {\bf #1} (#2) #3}

\nc{\cqg}[3]{Class. Quant. Grav. {\bf #1} (#2) #3}

\nc{\jhep}[3]{J. High Energy Phys. {\bf #1} (#2) #3}

\nc{\hep}[1]{{\tt hep-th/{#1}}}

%%%%%%%%%%%%%%%%%%%%%%%%%%%%%%%%%%%%%%%%%%%%%%%%%%%%%%%%%%%

\end{document}